\documentclass[sigconf]{acmart}
\AtBeginDocument{%
  }

\setcopyright{acmlicensed}
\copyrightyear{2025}
\acmYear{2026}
\acmConference[AI-SQE]{The 1st International Workshop on AI for Software Quality Evaluation: Judgment, Metrics, Benchmarks, and Beyond}{April 14, 2026}{Rio de Janeiro, Brazil}
\usepackage{multicol}
\usepackage{multirow}
\usepackage{enumitem}
\usepackage{graphicx} % Required for inserting images
\usepackage{booktabs}
\usepackage{rotating}
\usepackage{hyperref}
\usepackage{algorithm}
\usepackage{algpseudocode}
\usepackage{amsmath}
\usepackage{relsize}
\usepackage{tikz}
\usepackage{float}
\usepackage{makecell}
\usepackage{placeins}
\usetikzlibrary{shapes.geometric, arrows}
\usepackage{url}
\usepackage{hyperref}  % makes clickable links in references
\usetikzlibrary{arrows.meta,positioning,fit,calc}
\usepackage{xcolor}
\usepackage{listings}

\definecolor{PacBlue}{HTML}{1F4E79}   % Generation
\definecolor{PacTeal}{HTML}{1B998B}   % Evaluation
\definecolor{PacGreen}{HTML}{2F9E44}  % Inference
\definecolor{PacGray}{HTML}{4D4F53}   % Repos/Artifacts outlines
\definecolor{PacLite}{HTML}{F4F6F8}   % Subtle light fill

\begin{document}

\title{PACIFIC: a framework for generating benchmarks to check Precise Automatically Checked Instruction Following In Code}

\author{Itay Dreyfuss}
\author{Antonio Abu Nassar}
\author{Samuel Ackerman}
\author{Axel Ben David}
\author{Eitan Farchi}
\author{Rami Katan}
\author{Orna Raz}
\author{Marcel Zalmanovici}

\affiliation{%
  \institution{IBM Research}
  \city{Haifa}
  \country{Israel}
}
\email{{Itay.Dreyfuss, Antonio.Abu.Nassar, Samuel.Ackerman, Axel.Bendavid}@ibm.com}
\email{{Rami.Katan, farchi, ornar, marcel}@il.ibm.com}

\renewcommand{\shortauthors}{Dreyfuss, et al.}

\begin{abstract}
Large Language Model (LLM)-based code assistants have emerged as a powerful application of generative AI, demonstrating impressive capabilities in code generation and comprehension. A key requirement for these systems is their ability to accurately follow user instructions. We present Precise Automatically Checked Instruction Following In Code (PACIFIC), a novel framework designed to automatically generate benchmarks that rigorously assess sequential instruction-following and code dry-running capabilities in LLMs, while allowing control over benchmark difficulty.
PACIFIC produces benchmark variants with clearly defined expected outputs, enabling straightforward and reliable evaluation through simple output comparisons. 
% In contrast to existing approaches that often rely on tool usage or agentic behavior, our work isolates and evaluates the LLM’s intrinsic ability to simulate code execution mentally (dry running) and to follow instructions. 
In contrast to existing approaches that often rely on tool usage or agentic behavior, our work isolates and evaluates the LLM’s intrinsic ability to reason through code behavior step-by-step without execution (dry running) and to follow instructions.
Furthermore, our framework mitigates training data contamination by facilitating effortless generation of novel benchmark variations.
We validate our framework by generating a suite of benchmarks spanning a range of difficulty levels and evaluating multiple state-of-the-art LLMs. Our results demonstrate that PACIFIC can produce increasingly challenging benchmarks that effectively differentiate instruction-following  and dry running capabilities, even among advanced models. Overall, our framework offers a scalable, contamination-resilient methodology for assessing core competencies of LLMs in code-related tasks.
\end{abstract}

\begin{CCSXML}
<ccs2012>
   <concept>
       <concept_id>10011007.10011074.10011099.10011693</concept_id>
       <concept_desc>Software and its engineering~Empirical software validation</concept_desc>
       <concept_significance>500</concept_significance>
       </concept>
 </ccs2012>
\end{CCSXML}
\ccsdesc[500]{Software and its engineering~Empirical software validation}

\keywords{Code instruction following benchmark, automated expected results generation, controlled level of benchmark difficulty, automated benchmark generation}
% %% A "teaser" image appears between the author and affiliation
% %% information and the body of the document, and typically spans the

\received{October 22, 2025}
\received[accepted]{November 23, 2025}

\maketitle
\begin{center}
\small \emph{Accepted at AI-SQE 2026.}
\end{center}
\vspace{1em}

\section{Introduction 
\label{sec:introduction}}

Large Language Models (LLMs) have demonstrated remarkable capabilities in natural language understanding and generation, leading to the development of highly effective code assistants. These systems rely heavily on their ability to follow user instructions accurately, a trait that is central to their utility in real-world software engineering tasks. While high-quality benchmarks exist for evaluating instruction-following in natural language domains, such as IFEval\cite{zhou2023ifeval}, there remains a major gap in the availability of benchmarks tailored to code-related instruction following. Existing benchmarks in this space often depend on complex evaluation methods, such as using LLMs themselves as judges, and typically lack deterministic metrics and a focus on code dry-running.

To address these limitations, we present PACIFIC, a framework designed to automatically generate benchmarks that evaluate both sequential instruction-following and code dry-running capabilities of LLMs. The framework emphasizes the ability of LLMs to simulate code execution (i.e. dry-running) without relying on external tools or agentic mechanisms. 

The framework operates on a set of instructions that are used as building blocks. Each instruction describes actions to perform on the output of a given algorithm when dry-running it with a given input. Below are a couple of examples of such instructions:

\begin{enumerate}
    \item \textbf{next\_\allowbreak perfect\_\allowbreak square}: take the previous number and output the first perfect square that comes after it. E.g. if input is 4, output the next perfect square which is 9.
    \item \textbf{shift\_\allowbreak back}: subtract 1 from each character of the previous output; a becomes z. E.g., if input is IBM, output HAL.
\end{enumerate}

The full list of instructions in natural language can be found in appendix \ref{append:instructions}. The same instructions are given as code. For example, a corresponding C++ implementation for \textbf{next\_\allowbreak perfect\_\allowbreak square} is:

\begin{lstlisting}[language=C++, basicstyle=\ttfamily\small, frame=single]
int next_perfect_square(int n) {
    double sqrt_n = std::sqrt(n);
    int next_int = std::ceil(sqrt_n);
    if (next_int * next_int == n)
        next_int += 1;
    return next_int * next_int;}
\end{lstlisting}

These instructions can be concatenated to form complex multi-step tasks, either as a single prompt or in a chat-based format, with the number of concatenations controlled by a user-defined parameter.
Multiple instructions can then be composed into a pipeline, where each instruction receives as input the output produced by the preceding one.

We note that the instructions which are the framework's building blocks are made to be easily dry-run by a first year computer science major, while being complex for an LLM as demonstrated in our experiment. This is inspired by the ARC benchmark approach \cite{DBLP:journals/corr/abs-1911-01547}, but differs in the implementation and domain.

Each benchmark generated by PACIFIC can be unique, mitigating the risk of training data contamination. The expected results are computed by executing reference implementations of the described tasks, enabling automatic and deterministic evaluation via simple output comparison. We demonstrate the effectiveness of the framework by generating a series of benchmarks, showing that even state-of-the-art LLMs struggle with certain instruction combinations. 

Our results highlight the potential of our framework to produce challenging, scalable, and contamination-resilient benchmarks for assessing core LLM competencies in code instruction following and dry-running.
The main contributions of our paper are:
\begin{itemize}
    \item We introduce PACIFIC, a novel framework for automatically generating benchmarks that evaluate both instruction-following and code dry-running capabilities of LLMs. 
    The framework supports benchmark generation with customizable instruction concatenation, enabling scalable creation of diverse and contamination-resilient evaluation sets.
    \item PACIFIC benchmarks allow deterministic evaluation, relying on simple output comparison against expected results generated via reference code, without requiring tool use or LLM-as-a-judge paradigms.
    \item We introduce an approach to control sample difficulty and demonstrate that our defined difficulty dimensions correlate negatively with LLM performance. The difficulty dimensions are:
    \begin{enumerate}
        \item \textbf{LLM input:} the number of instructions in each sample (prompt).
        \item \textbf{LLM expected output:} the expected length of each sample's output.
    \end{enumerate}
\end{itemize}
We demonstrate the framework’s effectiveness by evaluating several LLMs on PACIFIC generated benchmarks of increasing difficulty. The results assist in differentiating among the different evaluated models. We notice that by controlling difficulty our framework can generate benchmarks that could also challenge the ever improving LLMs. 
Our results highlight the importance of code instruction following and dry-running ability as core competencies for LLM-based code assistants, which are not adequately captured by existing benchmarks.

 Section~\ref{sec:criteria} lists the criteria underlying our framework. Section~\ref{sec:pacifics_design} details the framework's design. Section~\ref{sec:experiment} provides an example of benchmarks generated by PACIFIC. Section~\ref{sec:results} lists the results of evaluating LLMs with the generated benchmarks. Section ~\ref{sec:related} discusses related work. Section ~\ref{sec:validity} lists the possible threats to validity. Section~\ref{sec:conclusion} concludes.

\textit{Terminology Note:} Throughout this paper, we use the terms \emph{complexity} and \emph{difficulty} interchangeably. Also, the terms \emph{prompt} and \emph{sample} are treated as equivalent in the context of our discussion.

\section{Criteria
\label{sec:criteria}}

To ensure robustness, adaptability, and long-term utility, we define the following core principles for our benchmark generation framework:

\begin{enumerate}
    \item \textbf{Simple and Deterministic Evaluation.} 
    The evaluation process must rely on deterministic and transparent metrics rather than LLM-based evaluators. Dependence on another model introduces variability and bias, as well as constraints tied to the evaluator’s capabilities. By employing simple, rule-based metrics, we guarantee evaluations that are fast, reproducible, and unambiguous.
    \item \textbf{Difficulty Control.} 
    The framework must provide explicit mechanisms for controlling the difficulty of the generated benchmarks. This capability enables the systematic creation of tasks ranging from easy to highly challenging, ensuring that the framework can assess both current state-of-the-art models and future frontier systems.
    \item \textbf{Scalability and Contamination Resistance.} 
    To produce meaningful results, the framework must support large-scale and diverse benchmark generation. Furthermore, it should allow rapid creation of alternative benchmark versions to mitigate training data contamination risks. If a benchmark version becomes part of a model’s training data, the framework should be able to generate an equivalent, unseen variant for unbiased evaluation.
    \item \textbf{Modularity and Extensibility.} 
    The framework should adopt a modular architecture that facilitates easy modification and extension. This design ensures that components can be added, replaced, or adapted with minimal effort, allowing the framework to evolve alongside emerging research needs and application domains.
\end{enumerate}

\section{PACIFIC's Design
\label{sec:pacifics_design}}

% In the body:
\begin{figure}[htbp]
    \centering
    \includegraphics[width=0.9\linewidth]{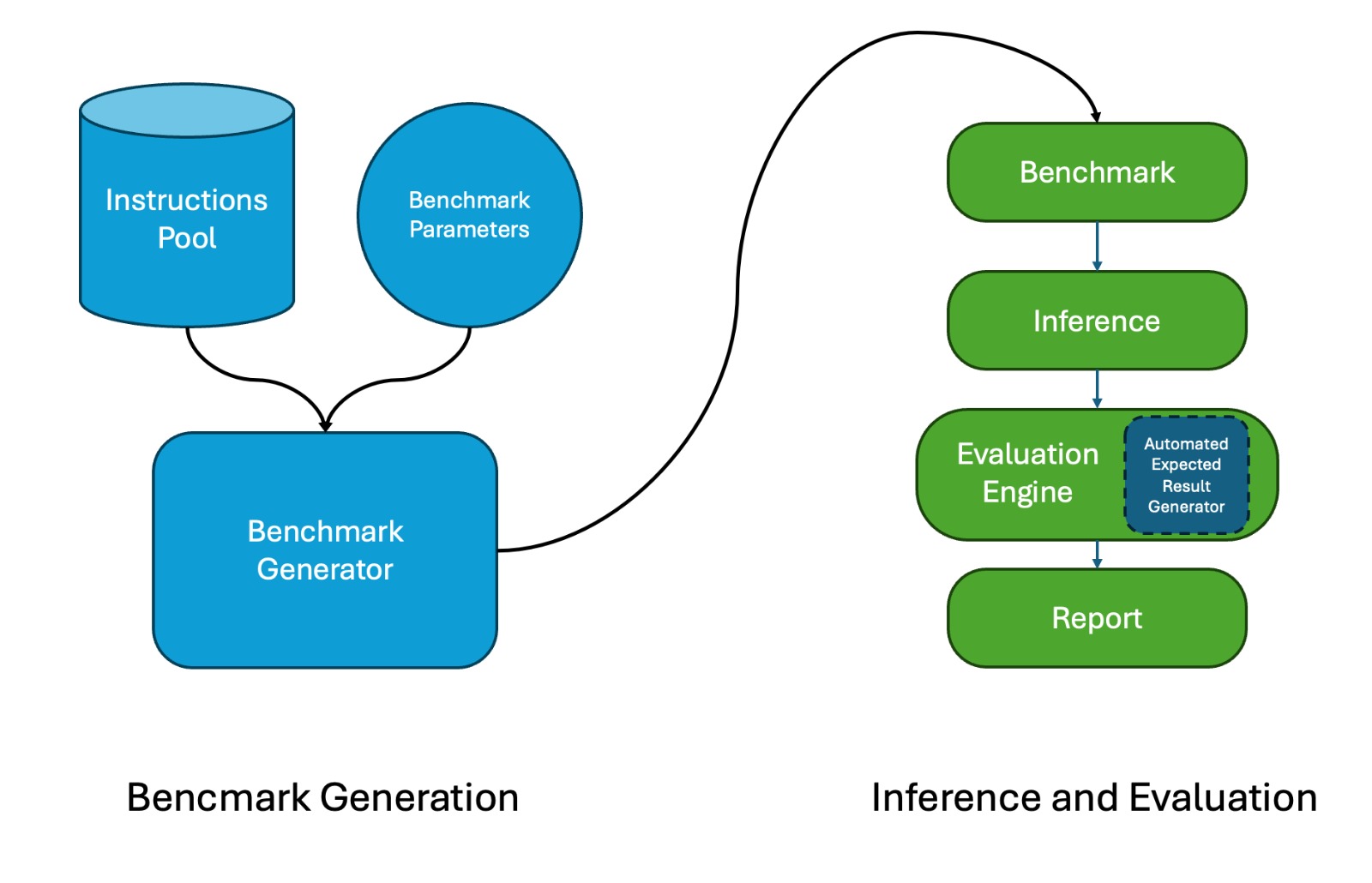}
    \caption{\textbf{PACIFIC: High-level Architecture.} Instruction Pool and Benchmark generation parameters feed the Benchmark Generator, producing a Benchmark. Inference runs on the benchmark, and the Evaluation Engine outputs metrics in the form of a report file.}
    \label{fig:pacific-architecture}
    \Description{PACIFIC high-level architecture diagram showing the main system components.}
\end{figure}

Our framework is designed to adhere to the core criteria outlined in Section~\ref{sec:criteria}. Below, we describe PACIFIC's architecture, and how each criterion is operationalized and implemented within the framework.
Each benchmark generated by PACIFIC is a set of many \textit{samples}, where each sample includes:
\begin{enumerate}
    \item \textbf{An initial input} --- either a number or a string.
    \item \textbf{A sequence of instructions} --- operations to be applied to the input. See the current set of initial instructions in Appendix\ref{append:instructions}.
\end{enumerate}
The model under evaluation is tasked with executing the instructions on the given input and producing the final output in the specified format.

\textbf{Instruction Structure.}
Instructions serve as the fundamental building blocks of PACIFIC. Each instruction is a function that accepts either a number or a string and returns a value of one of those types. Currently, instructions are implemented in multiple programming languages, including Python, Java, and C++. We drew inspiration for making instructions the building blocks of our framework from CodeIF~\cite{yan2025codeif}.

To illustrate the diversity of instruction types, we provide two representative examples: one mapping a number to a number, and another mapping a number to a string. While the examples below are expressed in natural language, the framework stores their code implementations in all supported programming languages.

\begin{itemize}
    \item Instruction 1:
    \begin{quote}
        Given the previous number, output the smallest number bigger than the previous answer that is a prime number.
    \end{quote}
    \item Instruction 2:
    \begin{quote}
        Take the previous number n, and output the day of the week with index (n mod 7) - sunday is the first day and its index is 0, in lower caps.
    \end{quote}
\end{itemize}
These examples illustrate two of the four possible input–output type combinations: number-to-number and number-to-string. The framework supports all four combinations.

\subsection{Benchmark Generation Parameters}
To generate a benchmark, the following parameters must be specified:
\begin{itemize}
    \item \textbf{Number of Instructions:} The number of instructions per sample. Increasing this value results in longer and more complex outputs.
    \item \textbf{Target Output Length:} The desired length of the final output. For strings, this is measured in characters; for numbers, in bits. Larger targets increase task difficulty.
    \item \textbf{Samples per Programming Language:} The number of unique samples to generate for each supported language.
    \item \textbf{Mode:} Either \textit{Prompt} or \textit{Chat}. In \textit{Prompt} mode, the entire sample is presented as a single prompt; in \textit{Chat} mode, instructions are provided sequentially in a user--assistant dialogue format.
    \item \textbf{Random Seed:} Ensures reproducibility of random choices. Different seeds produce distinct benchmarks.
    \item \textbf{Instruction Type:} Either \textit{Code} (instructions as code snippets) or \textit{NL} (instructions expressed in natural language).
\end{itemize}

\subsection{Benchmark Generation Process}
\label{subsec:benchmark_generation_process}
For each sample instructions are chosen from the instruction pool under two constraints:
\begin{enumerate}
    \item \textbf{Type Consistency:} The input type of each instruction must match the output type of the previous instruction.
    This constraint ensures that no type mismatches occur.
    \item \textbf{Length Adjustment:} The system dynamically selects instructions to control output size, using length-aware categories to meet the target length.
    This mechanism enables fine-grained control over output length during benchmark construction.
\end{enumerate}
Beyond these constraints, instruction selection is random and fully reproducible via the specified random seed. This process is repeated until the required number of samples is generated, completing the benchmark.

\begin{figure}[htbp]
    \centering
    \includegraphics[width=0.9\linewidth]{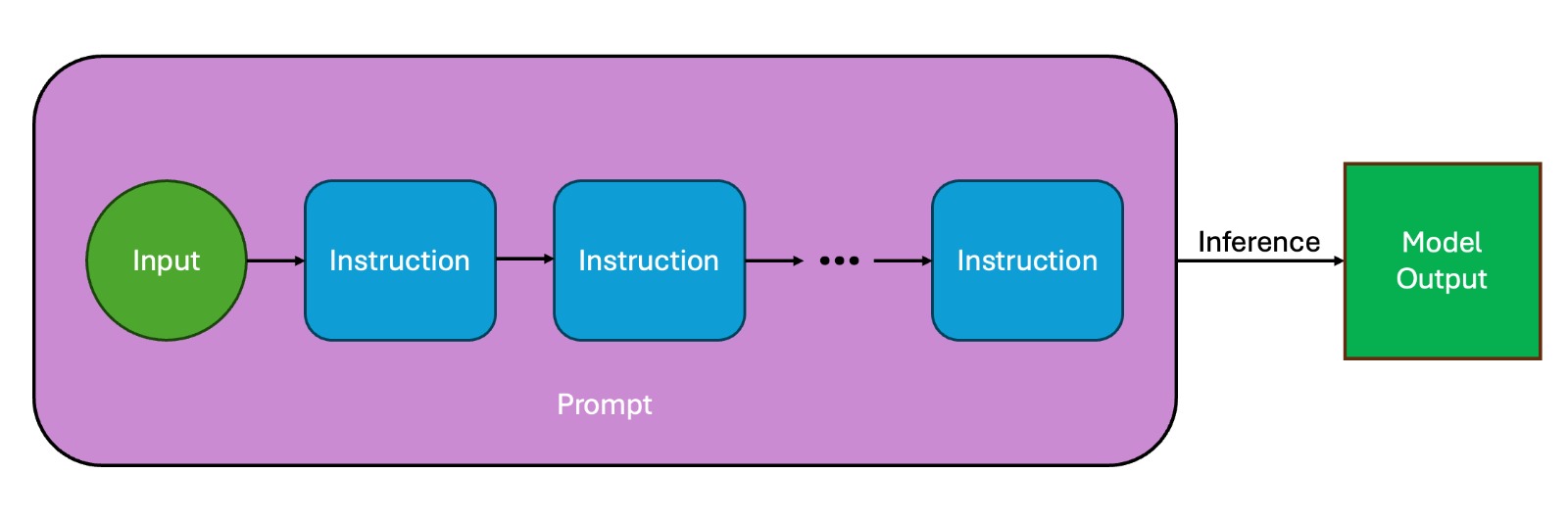}
    \caption{\textbf{Benchmark Sample Structure.} A sample consists of an initial input followed by a sequence of instructions forming the prompt, which is processed by the model to generate the output.}
    \label{fig:benchmark-sample}
    \Description{Benchmark sample structure diagram.}
\end{figure}

Figure~\ref{fig:benchmark-sample} illustrates the inference pipeline: the input is combined with a chain of instructions to form the prompt, which is then passed to the model for inference, producing the final output.
We drew inspiration from CruxEval~\cite{gu2024cruxeval} on this process, mainly on how to keep both the benchmark generation and the metrics simple.

\subsection{How PACIFIC Meets the Criteria}
We now describe how PACIFIC’s architecture satisfies each of the framework’s design principles.

\subsubsection{Simple and Deterministic Evaluation.}
\label{subsec:evaluation}

To ensure transparency, reproducibility, and efficiency, PACIFIC employs a fully rule-based evaluation pipeline. Inspired by the principles introduced in IF-Eval~\cite{zhou2023ifeval}, we emphasize determinism as a core property of our evaluation methodology. This approach avoids reliance on LLM-based evaluators, which can introduce bias, variability, and significant computational overhead.

\textbf{Evaluation Process.} For each response, our framework parses the output to extract intermediate results after each instruction.

\textbf{Correctness Check:} The framework validates each instruction by comparing the output against the expected result. Inspired by the metric design proposed in~\cite{zhou2023ifeval}, we compute two metrics:
\begin{itemize}
    \item \textbf{Prompt-Level Accuracy:} The fraction of samples in which all instructions were followed correctly.
    \item \textbf{Instruction-Level Accuracy:} The average fraction of correctly executed instructions across all samples.
\end{itemize}

\textbf{Error Handling:} PACIFIC classifies incomplete or invalid outputs into predefined categories, such as insufficient answers or type mismatches, while valid samples proceed to evaluation.

\textbf{Advantages.} These metrics are deterministic, ensuring consistent results across runs. They also require negligible computational resources compared to LLM-based evaluators, which would effectively double the benchmark cost in time and compute.

\subsubsection{Difficulty Control.}
PACIFIC provides explicit mechanisms for controlling task difficulty through a set of adjustable parameters. This design enables systematic scaling of benchmark difficulty, ensuring applicability across a wide range of models, from current models to future state-of-the-art ones.

\textbf{Primary Difficulty Regulators.} Two key parameters govern the difficulty of generated tasks:
\begin{itemize}
    \item \textbf{Number of Instructions:} Serves as a quantitative difficulty regulator. Increasing the number of instructions requires the model to perform longer reasoning chains. As illustrated in Figure~\ref{fig:benchmark-sample}, more instructions extend the prompt, which increases error likelihood as seen in Figure~\ref{fig:complexity-performance}.
    \item \textbf{Targeted Output Length:} Acts as a qualitative difficulty regulator. Larger output lengths introduce additional difficulty by requiring the model to handle more extensive transformations and maintain consistency across longer sequences. In the pipeline shown in Figure~\ref{fig:benchmark-sample}, this means each instruction must produce more complex intermediate results, stressing both reasoning and generation capabilities.
    We control the targeted output length using the length adjustment constraint introduced in Section~\ref{subsec:benchmark_generation_process}. 
    To achieve this, instructions are categorized based on their effect on output size: some increase length, others reduce it. By monitoring intermediate expected lengths during benchmark construction, the system selects subsequent instructions to keep the overall output length close to the targeted length.
\end{itemize}

\textbf{Adaptability and Relevance.} By tuning these parameters, the framework allows to select difficulty levels appropriate for the model under evaluation or to probe the model’s capability limits. This adaptability ensures that the benchmark remains relevant over time, even as models improve, and supports controlled experiments for analyzing performance under varying levels of complexity.

\subsubsection{Scalability and Contamination Resistance.}
Our framework is engineered to support large-scale, diverse benchmark generation while enabling rapid creation of equivalent, unseen variants to mitigate contamination risks.

\textbf{Contamination Resistance.} The framework enables regeneration of semantically equivalent yet unseen benchmark variants via multiple, orthogonal mechanisms:
\begin{itemize}
    \item \textit{Seed-based resampling.} Changing the random seed re-samples inputs and instruction sequences subject to the same constraints, preserving target difficulty while altering content.
    \item \textit{Representation diversity.} The same instruction sequence can be rendered as \textit{Code} (Python/Java/C++) or \textit{NL} (natural-language) instructions, and delivered in either \textit{Prompt} or \textit{Chat} mode. These variants maintain the underlying computational graph but change surface form and discourse structure, reducing memorization and template overfitting.
\end{itemize}

Together, these design choices support generating large benchmarks while enabling rapid, difficulty-matched, unseen variants for robust contamination mitigation.

\subsubsection{Modularity and Extensibility.}
PACIFIC adopts a modular architecture that separates components into distinct layers, including:
\begin{itemize}
    \item \textbf{Instruction Pool:} A repository of reusable instructions implemented in multiple programming languages.
    \item \textbf{Benchmark Generator:} Responsible for sampling inputs, assembling instruction sequences, and enforcing complexity constraints.
    \item \textbf{Evaluation Engine:} Implements deterministic, rule-based metrics for correctness and error categorization.
\end{itemize}
This modular design ensures that components can be independently added, replaced, or extended with minimal engineering effort, enabling the framework to evolve alongside emerging research needs and application domains.

\textbf{Adding Custom Instructions.} PACIFIC provides an automated and user-friendly procedure for extending the instruction pool with new custom instructions.
The procedure supports batch mode, allowing multiple instructions to be added in a single execution, significantly reducing manual overhead. Once an instruction is correctly added to the pool, it becomes immediately available for:
\begin{itemize}
    \item \textbf{Benchmark Generation:} All future benchmarks incorporate the new instruction without additional configuration.
    \item \textbf{Evaluation:} The evaluation engine automatically recognizes and validates outputs for the new instruction, requiring no extra integration steps.
\end{itemize}
This automated extensibility mechanism ensures that our framework can rapidly adapt to new research scenarios, incorporate domain-specific operations, and maintain long-term relevance with minimal maintenance cost.

\section{Experiment Setup}
\label{sec:experiment}

To evaluate PACIFIC, we designed an experiment comprising of 15 distinct generated benchmarks. Each benchmark was generated by varying two difficulty parameters: the number of instructions $I = \{3, 5, 8, 10, 15\}$ and the targeted output length $L = \{3, 5, 10\}$. These choices yield $5 \times 3 = 15$ unique configurations.

For all benchmarks, we fixed the following parameters to ensure comparability:
\begin{itemize}
    \item \textbf{Sample Size:} Each benchmark contained 99 samples (33 per programming language across three languages -- Python, Java, and C++).
    \item \textbf{Prompt Mode:} We adopted prompt-based evaluation rather than multi-turn chat to reduce time and token cost.
    \item \textbf{Random Seed:} A fixed seed was used for reproducibility.
    \item \textbf{Instruction Type:} All instructions were code-implemented, excluding natural language formulations.
\end{itemize}

We evaluated the generated benchmarks on a diverse set of LLMs, including:
\begin{itemize}
    \item \texttt{Claude 3.5 Haiku}~\cite{hf_claude35_haiku}
    \item \texttt{Claude 4 Sonnet}~\cite{hf_claude4_sonnet}
    \item \texttt{gpt-oss-120b}~\cite{hf_gptoss120b}
    \item \texttt{Llama-4-Maverick-17B-128E-Instruct}~\cite{hf_llama4_maverick}
    \item \texttt{Qwen3-235B-A22B-Instruct-2507}~\cite{hf_qwen3_235b}
    \item \texttt{Qwen3-Coder-480B-A35B-Instruct-FP8}~\cite{hf_qwen3_coder_480b}
\end{itemize}
% \texttt{Claude 4.1 Opus} was initially considered; however, it was excluded from the final evaluation due to its significantly higher token cost and inferior performance compared to GPT-OSS 120B.

Our design choices aim to demonstrate three core contributions:
\begin{enumerate}
    \item The inherent difficulty of dry-running code and following code-specific instructions for current LLMs.
    \item The distinction between this capability and general NL instruction-following or code generation tasks.
    \item A novel mechanism for controlling sample difficulty via adjustable parameters (instruction count and output length).
\end{enumerate}

\section{Experiment Results}
\label{sec:results}

In this section, we present the outcomes of evaluating PACIFIC across the 15 generated benchmarks and multiple LLMs. The results demonstrate the effectiveness of the proposed framework in achieving its objectives.

\subsection{Overview of Findings}
The evaluation reveals three key insights:
\begin{enumerate}
    \item \textbf{Difficulty Control:} Varying the number of instructions and targeted output length successfully modulates task difficulty, confirming the viability of our mechanism for controlled benchmark generation. See Section~\ref{subsec:complexity-control}.
    \item \textbf{Distinct Nature of the Task:} Model rankings differ from those observed in NL instruction-following or code generation tasks, reinforcing the uniqueness of the evaluated capability. See Section~\ref{subsec:nature-task}.
    \item \textbf{Skill Gap in Dry-Running Code:} All tested models exhibit significant performance degradation as difficulty increases, highlighting that dry-running code and precise instruction-following remain challenging for state-of-the-art LLMs. See Sections~\ref{subsec:model-comparison}.
\end{enumerate}
While these results expose a clear skill gap, it is important to note that such tasks are easier to solve when models are augmented with external tools or API-calling capabilities, as they can execute the code rather than reason through it. However, our objective was to assess the intrinsic reasoning for code dry-running and instruction following abilities of LLMs in isolation, rather than their tool usage and agentic implementation performance.

\subsubsection{Performance Across Difficulty Levels}
\label{subsec:complexity-control}

\begin{figure*}[htbp]
    \centering
    % Placeholder for complexity vs performance graph
    \includegraphics[width=\linewidth]{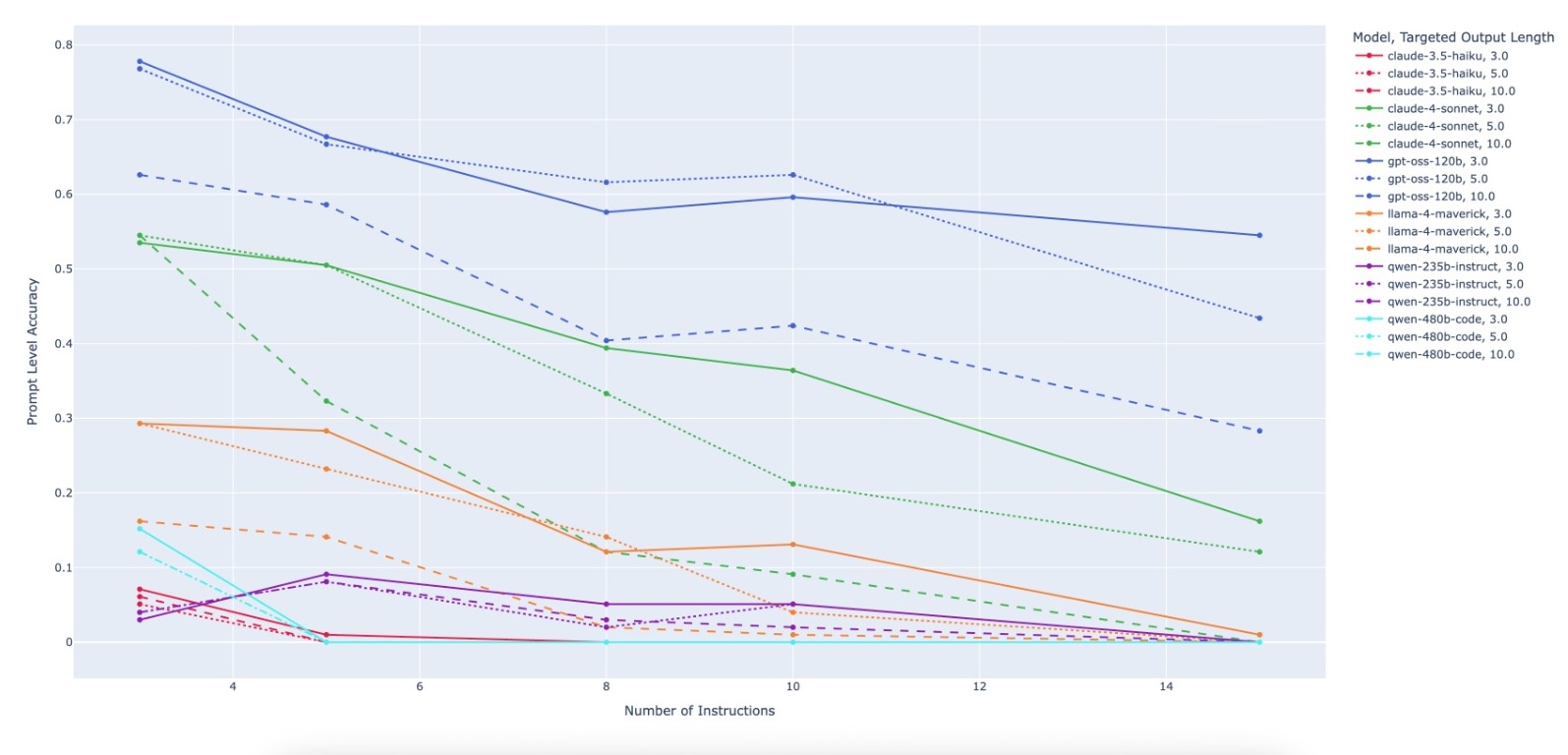}
    \caption{Prompt Level Accuracy across varying complexity levels. Higher instruction counts and output lengths correspond to increased task difficulty. Average across all programming languages.}
    \label{fig:complexity-performance}
    \Description{Prompt level accuracy graph}
\end{figure*}

Figure~\ref{fig:complexity-performance} illustrates Prompt-Level Accuracy (y-axis) as a function of benchmark difficulty, measured by the number of instructions (x-axis). Each color represents a different model, while line styles (solid, dashed, dotted) within the same color correspond to varying targeted output lengths (1.0, 3.0, 5.0, and 10.0). Thus, curves of the same color indicate the same model evaluated under different output-length constraints.

The results reveal a clear negative correlation between difficulty and correctness across all models. As the number of instructions increases, performance consistently declines. Furthermore, for a fixed instruction count, increasing the targeted output length (represented by different line styles within the same color) leads to additional degradation in accuracy. This trend suggests that both instruction count and output length impose compounding challenges on instruction-following capabilities.

All reported metrics are averaged across the three supported programming languages. While per-language statistics can be extracted, we do not present them here as they exhibited minimal variance in our experiments.

\subsubsection{Distinct Nature of the Task}
\label{subsec:nature-task}

As shown in Table~\ref{tab:performance-across-tasks}, performance rankings on PACIFIC's evaluated task diverge substantially from those observed in general instruction-following or code generation benchmarks. For example, \texttt{gpt-oss-120b}~\cite{hf_gptoss120b} ranks first in PACIFIC, yet only fourth in both instruction-following and coding tasks, indicating that its strengths are uniquely aligned with PACIFIC's requirements rather than conventional paradigms. Conversely, \texttt{Qwen3-235B-A22B-Instruct-2507}~\cite{hf_qwen3_235b}, which dominates coding (rank 1) and instruction-following (rank 1), falls to fourth place on PACIFIC, suggesting that proficiency in these established benchmarks does not guarantee success in our setting. In contrast, \texttt{claude-4-sonnet}~\cite{hf_claude4_sonnet} maintains relatively stable performance across all tasks (ranked 2 in PACIFIC and instruction-following, and 2 in coding), while \texttt{
Llama-4-Maverick-17B-128E-Instruct}~\cite{hf_llama4_maverick} exhibits strong PACIFIC performance (rank 3) but degrades sharply in coding (rank 6). 

These discrepancies underscore that PACIFIC evaluates a distinct capability, compositional instruction execution under difficulty constraints, rather than a mere variant of existing instruction-following or code-generation tasks.

\begin{table}[htbp]
  \centering
  \caption{Relative performance across different tasks. The PACIFIC column is the task used in our experiment, The Instruction Following and Coding are taken from LMArena's leaderboard~\cite{lmarena2025leaderboard}. The numbers represent the relative ranking. Lower is better.}
  \label{tab:performance-across-tasks}
  \begin{tabular}{l|c|c|c}
    \toprule
    \textbf{Model} & \textbf{PACIFIC} & \textbf{Instruction} & \textbf{Coding} \\
     &  & \textbf{Following} &  \\
    \midrule
    gpt-oss-120b        & 1 & 4 & 4 \\
    claude-4-sonnet     & 2 & 2 & 2 \\
    Llama-4-Maverick      & 3 & 5 & 6 \\
    Qwen3-235B-A22B       & 4 & 1 & 1 \\
    Qwen3-Coder-480B-A35B      & 5 & 3 & 3 \\
    claude-3.5-haiku    & 6 & 6 & 5 \\
    \bottomrule
  \end{tabular}
\end{table}

\subsubsection{Model Comparison}
\label{subsec:model-comparison}

\begin{figure}[htbp]
    \centering
    \includegraphics[width=0.9\linewidth]{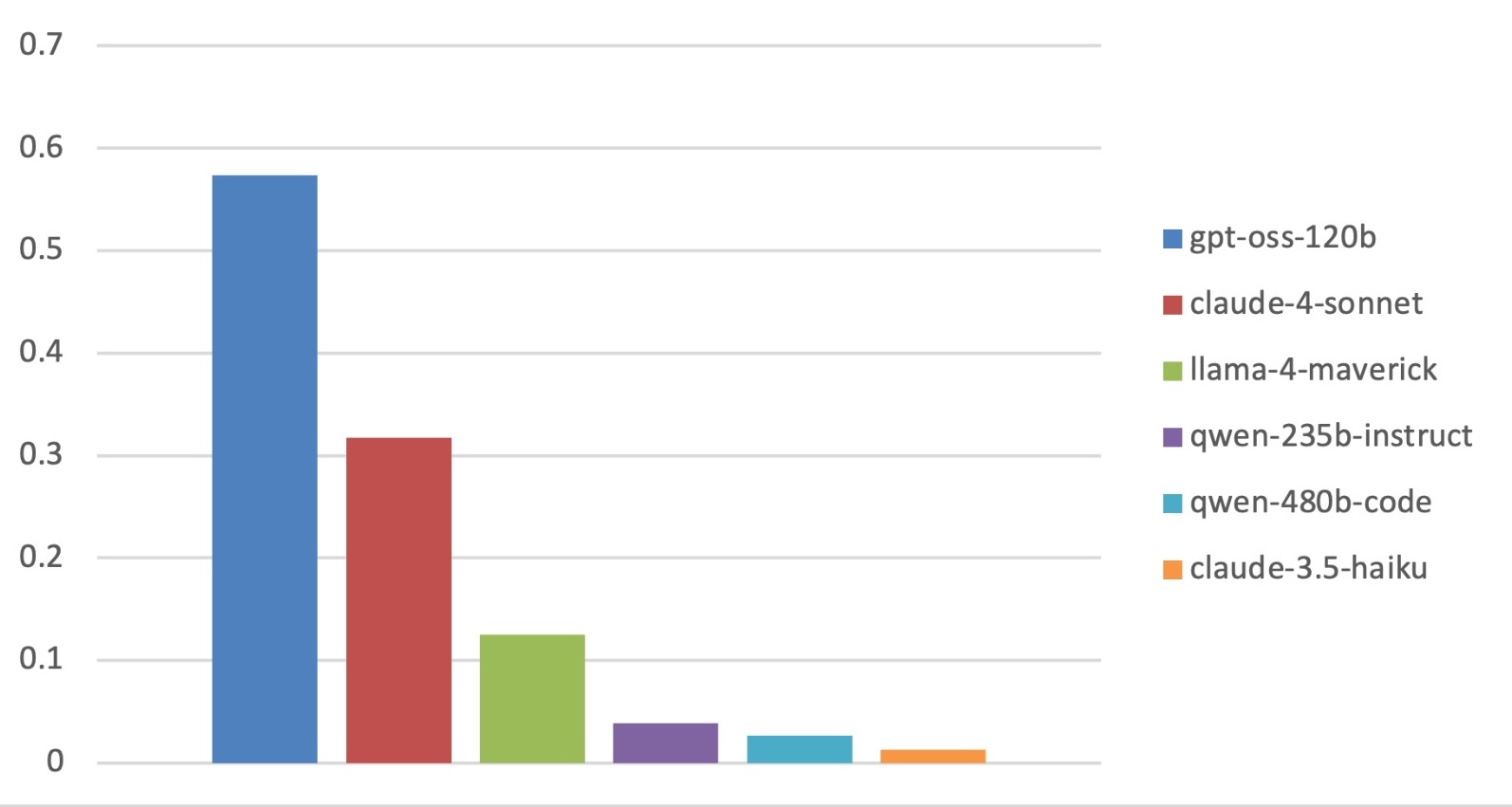}
    \caption{Aggregate performance of evaluated LLMs, measured as the mean Prompt Level Accuracy across all benchmarks.}
    \label{fig:model-comparison}
    \Description{Aggregate performance of evaluated LLMs}
\end{figure}

Figure~\ref{fig:model-comparison} shows the aggregate performance of all evaluated models, computed as the mean Prompt Level Accuracy across all benchmarks (including both easy and difficult tasks). This comparison reveals clear differences in capability: while larger models generally achieve higher accuracy than smaller ones, the variation across models highlights which architectures are currently better suited for dry-running code.

One can see, for example, that \texttt{claude-4-Sonnet}~\cite{hf_claude4_sonnet} achieved a poor score in this task. Though we measured high success rate for each individual instruction, examples of which can be seen in Figure~\ref{fig:instructions-samples}, the model rarely succeeds to answer all instructions correctly when given multiple instructions in a single prompt. The aggregated performance scores in Figure~\ref{fig:model-comparison} are computed via \textbf{Prompt Level Accuracy} which is a strict metric that requires all instructions in a given sample to be followed correctly. The metric \textbf{Instruction Level Accuracy} quantifies the percentage of correctly answered instructions but is not shown here, as the emphasis is on instruction following. Both metrics are defined in Section~\ref{subsec:evaluation}.

\begin{figure}[htbp]
    \centering
    \includegraphics[width=0.9\linewidth]{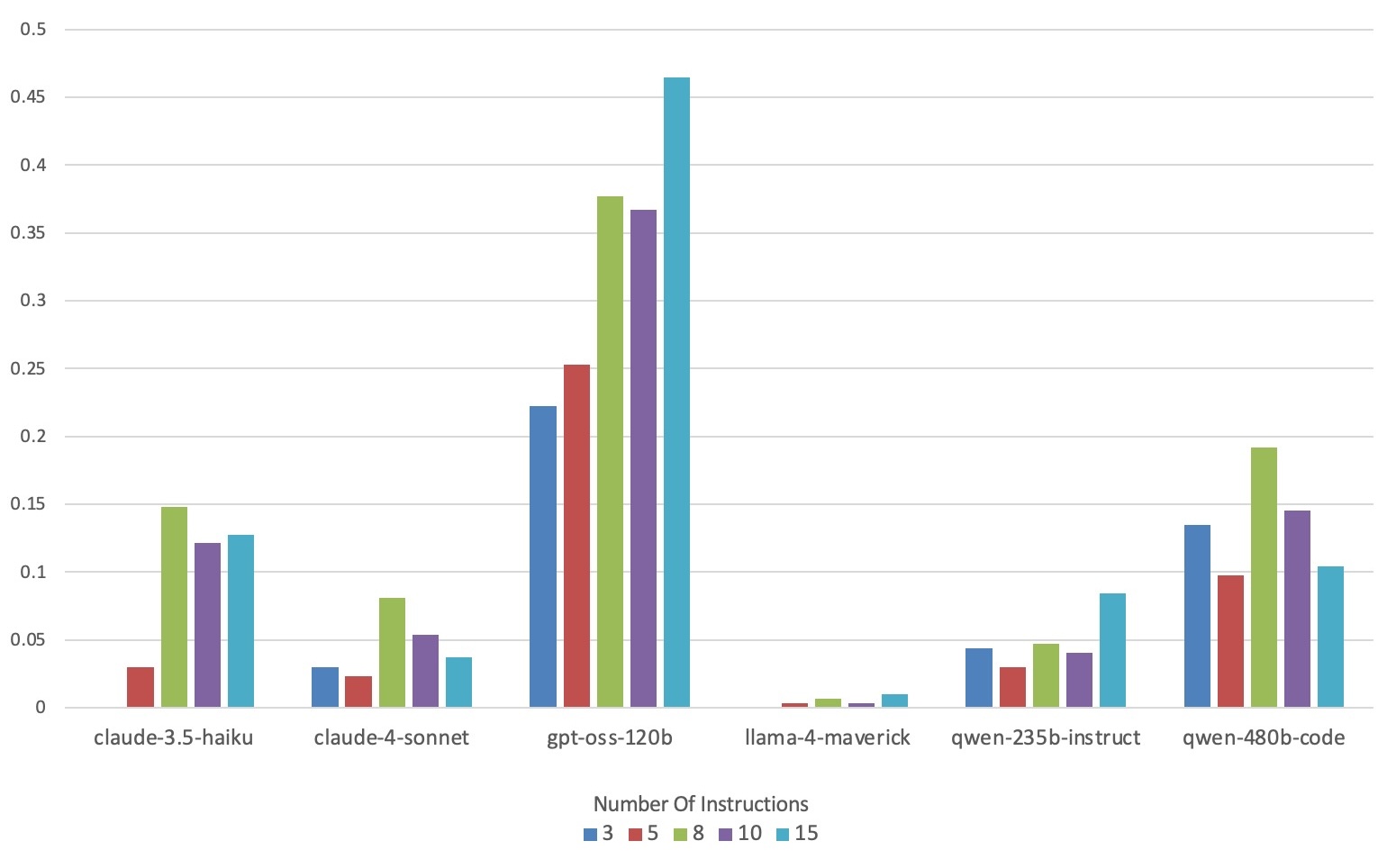}
    \caption{Aggregate missing answers rates across models and instruction counts. The higher the instruction count, the higher the error rate.}
    \label{fig:error-rates}
    \Description{Aggregate missing answers rates}
\end{figure}

\textbf{Error Rates}
Figure~\ref{fig:error-rates} illustrates the aggregated rates of missing answers across models for varying instruction counts. The results reveal a clear trend: error rates generally increase as the number of instructions grows, indicating that longer prompts pose greater challenges for accurate execution and formatting. While some models maintain low error rates even at higher instruction counts (e.g., \texttt{Llama-4-maverick}), others exhibit a steep rise (e.g., \texttt{gpt-oss-120b}~\cite{hf_gptoss120b}), highlighting differences in robustness and scalability among the evaluated models. 

It is important to note that \textit{Targeted Output Length} parameter exhibits minimal variance in error rates across its different values, suggesting that output length has negligible influence compared to instruction count in this context.

We observe that most errors produced by \texttt{gpt-oss-120b}~\cite{hf_gptoss120b} originate from its inability to comply with strict formatting requirements. In contrast, \texttt{Qwen3-235B-A22B-Instruct-2507}~\cite{hf_qwen3_235b} generally adheres to the expected output format but fails to deliver correct solutions, primarily due to the absence of effective reasoning strategies such as chain-of-thought prompting introduced in~\cite{wei2023chainofthoughtpromptingelicitsreasoning}.

\subsubsection{Instruction Samples}
\label{subsec:instruction-samples}

\begin{figure*}[htbp]
    \centering
    % Placeholder for complexity vs performance graph
    \includegraphics[width=\linewidth]{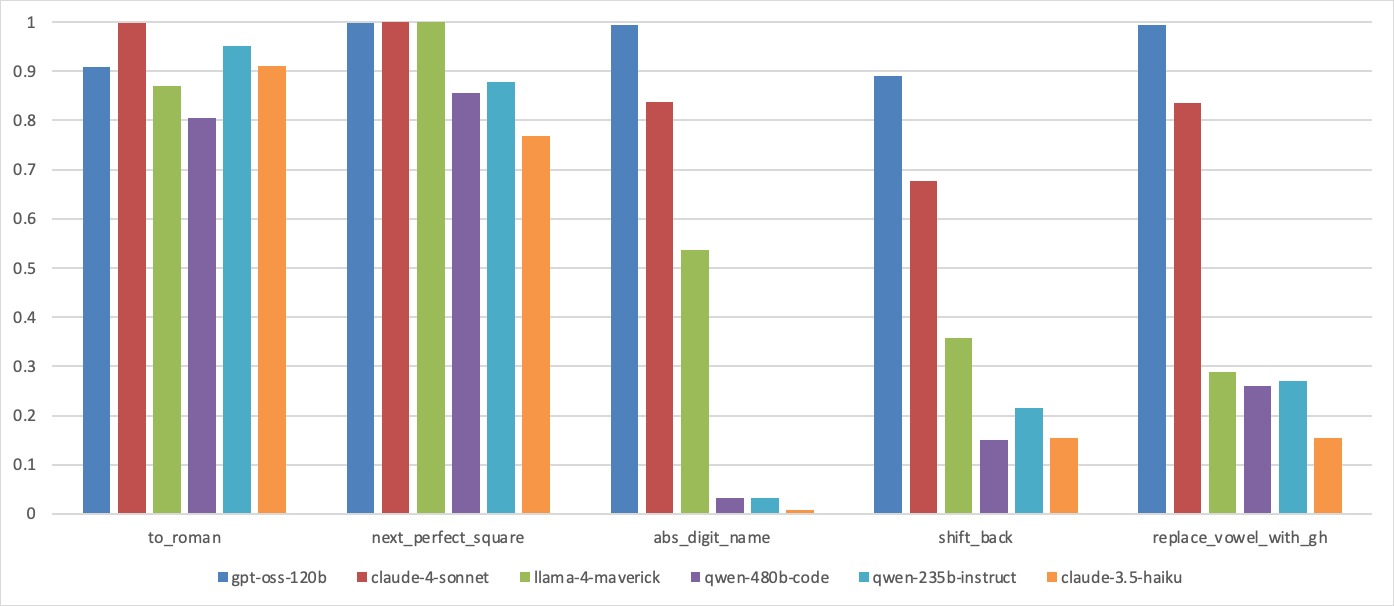}
    \caption{Accuracy across models for five representative instructions, grouped by complexity. The to\_roman and next\_perfect\_squared are considered easy, the other ones are the hardest in our current instruction set. Scores are aggregated by instruction count and targeted output length.}
    \label{fig:instructions-samples}
    \Description{Accuracy across models for five representative instructions}
\end{figure*}

Figure~\ref{fig:instructions-samples} presents a representative subset of instructions, with scores aggregated across instruction count and targeted output length for each model. The first two instructions, \textbf{to\_\allowbreak roman} and \textbf{next\_\allowbreak perfect\_\allowbreak square}, exemplify relatively simple tasks, as reflected by consistently high accuracy rates (approximately 85\%) across all models. We hypothesize that these tasks are easier for the models due to memorization effects rather than genuine reasoning.
In contrast, the last three instructions—\textbf{abs\_\allowbreak digit\_\allowbreak name}, \textbf{shift\_\allowbreak back}, and \textbf{replace\_\allowbreak vowels\_\allowbreak with\_\allowbreak gh}—are among the most challenging, despite being computationally easy for humans and requiring only linear time complexity. It is interesting to see that better models perform well on these instructions when they are not concatenated with other instructions, and yet that is not the case when multiple instructions are given.

A key insight is that \emph{compositional complexity}—combining multiple instructions within a single prompt—introduces a substantial performance bottleneck for current models. For instance, while \texttt{Claude-4-Sonnet} achieves over 65\% accuracy on both simple and complex instructions at the instruction level, its accuracy collapses to zero when these instructions are combined in the most difficult benchmark configuration.

\subsection{Summary}
These findings validate PACIFIC’s design goals:
\begin{itemize}
    \item The controlled variation of difficulty parameters enables systematic stress-testing of LLM capabilities. This flexibility ensures that the framework remains relevant even as future state-of-the-art models emerge, since benchmark difficulty can be scaled dynamically to match evolving capabilities.
    \item PACIFIC rankings diverge significantly from traditional instruction-following and coding benchmarks, demonstrating that our framework captures a distinct capability: compositional instruction execution under complexity constraints.
    \item The observed performance gap highlights an underexplored dimension of reasoning—code instruction following and dry-run-execution that is not adequately captured by existing benchmarks. Notably, almost all models scored \textbf{0\%} on the hardest benchmark configuration, and only one model reached \textbf{28\%}, which is still very low. This underscores the severity of the skill gap. Furthermore, our framework can effortlessly generate even harder benchmarks, making it a scalable tool for pushing the limits of LLM reasoning.
\end{itemize}
\section{Related work
\label{sec:related}}

Traditional benchmarks for evaluating language models on programming tasks, such as HumanEval~\cite{chen2021evaluating}, APPS~\cite{hendrycks2021apps}, and MBPP~\cite{austin2021program}, primarily measure \emph{functional correctness} in code generation. These datasets assess whether generated code passes a suite of unit tests, offering a reliable yet coarse measure of model competence. While effective for gauging program synthesis quality, such benchmarks do not explicitly test a model’s ability to follow fine-grained instructions or reason about intermediate execution steps. In contrast, PACIFIC focuses on \emph{instruction-level adherence} and \emph{reasoning consistency}, evaluating whether models correctly simulate execution or follow deterministic, step-by-step directives rather than merely producing functionally correct code.

Instruction-following has been extensively studied in the context of natural language tasks, with benchmarks such as IFEval~\cite{zhou2023ifeval} providing structured evaluations of model adherence to instructions. FollowBench~\cite{jiang2024followbench} builds on this idea by measuring multi-level, fine-grained compliance with explicit constraints, such as content, style, format, and examples. examples. Recently, newer studies have emerged focusing on evaluating LLMs' ability to follow sequences of instructions, such as the SIFO Instruction-following in code-related tasks remains underexplored, particularly in isolating reasoning without tool use or subjective judgment~\cite{chen2024sifobenchmarkinvestigatingsequential}. Our benchmark addresses this gap by emphasizing code logic reasoning and deterministic, stepwise evaluation.

CodeIF~\cite{yan2025codeif} introduces a benchmark for evaluating LLMs' ability to follow instructions in code generation tasks, including synthesis, debugging, and explanation. CodeIF-Bench~\cite{wang2025codeifbench} extends this framework to multi-turn interactions, assessing models across sequences of verifiable sub-instructions. Both benchmarks address instruction-following in generative contexts; however, they rely on tool-based or LLM-driven evaluation. Our approach emphasizes reasoning over code logic and dry-running, using simple deterministic metrics for assessment. 

CRUXEval~\cite{gu2024cruxeval} introduces an execution-based benchmarking methodology in which function outputs are generated by directly running the code, ensuring correctness and enabling rigorous evaluation of reasoning and understanding capabilities. Building on this principle, our framework adopts execution-driven output generation to verify model responses at the granularity of individual instructions rather than solely the final output.

EquiBench~\cite{wei2025equibench} assesses LLMs’ semantic understanding via equivalence checking between functionally identical programs, requiring paired programs and validation. In contrast, our benchmark uses natural language algorithm descriptions and evaluates step-by-step execution simulation, emphasizing instruction-following and comprehension.

AutoCodeBench~\cite{chou2025autocodebench} scales multilingual code benchmarks using LLMs, focusing on code generation and execution-based evaluation. In contrast, PACIFIC derives deterministic benchmarks from seed instructions, enabling automatic evaluation through simple output comparison.

InFoBench~\cite{qin-etal-2024-infobench} introduces the DRFR metric for decomposing instruction following performance. Our framework decomposes instruction following by construction, thus eliminating the need for such a complex metric.
\section{Threats to validity}\label{sec:validity}
The benchmarks generated by PACIFIC for the the experimental evaluation may be influenced by the seed instructions or by the initial algorithms to evaluate. We have not observed such an influence in our experiments, but it would be good to better understand what affects difficulty.

The difficulty of a PACIFIC benchmark may be affected by additional factors beyond the number of instructions and output size. Moreover, there may be combined affects of the two factors that we control (number of instructions and output size).

Chat mode in a PACIFIC benchmark may behave differently than the single prompt mode. We have anecdotally checked this and saw no difference, but a thorough investigation is needed. 

The initial version of PACIFIC consists of only three programming languages: Python, Java, and C++. Other, less popular, programming languages may behave differently. PACIFIC provides an easy procedure to add additional programming languages.

Contamination resistance, a core design objective of PACIFIC, has not yet been empirically validated across all scenarios. While the framework’s construction is intended to minimize contamination risk, confirming this property through systematic evaluation remains an important direction for future work.

The phrasing of PACIFIC’s prompts and the requested output format may influence LLM performance. Although the framework enforces instruction-following for code tasks, certain models may be less trained on the specific output format. We note that the requested format is intentionally simple; failure to follow it would indicate a fundamental limitation in instruction adherence.

PACIFIC generates benchmarks that target the evaluation of LLMs without tool usage. These benchmarks would be less difficult if tool usage is enabled, for example in an agentic implementation. 
\section{Conclusion}
\label{sec:conclusion}

In this work, we introduced PACIFIC, a principled framework for generating benchmarks that evaluate instruction-following and code dry-running capabilities of LLMs. Our framework addresses key limitations of existing benchmarks by providing deterministic, rule-based evaluation metrics that ensure transparency, reproducibility, and efficiency without relying on LLM-based judges. 

The framework incorporates explicit difficulty control through adjustable parameters such as the number of instructions and target output length, enabling systematic scaling of task difficulty to match diverse model capabilities. Furthermore, PACIFIC supports contamination-resistant benchmark generation via seed-based resampling and representation diversity, allowing rapid creation of semantically equivalent yet unseen variants. 

Its modular architecture facilitates extensibility, enabling researchers to easily integrate new instructions and adapt the benchmark to evolving domains. Experimental results demonstrate that PACIFIC can produce challenging tasks even for state-of-the-art models, highlighting persistent gaps in instruction-following and dry-running reasoning. 

Overall, PACIFIC provides a scalable, adaptable, and contamination-resilient solution for evaluating core competencies of LLM-based code assistants, paving the way for more rigorous and transparent assessment of future systems.

\bibliographystyle{ACM-Reference-Format}
\bibliography{main}


\clearpage
\appendix
\section{Appendix}
\label{appendix:appendix}

% Part 1: Instruction Set
\subsection{Current Set of Natural Language Instructions} \label{append:instructions}
\FloatBarrier
Below we list the current set of verbal instructions from which the PACIFIC benchmarks are constructed. This set is evolving. This is an inherent part of PACIFIC's design principles described in Section \ref{sec:pacifics_design}.
\begin{enumerate}
    \item Take the previous number and output the amount of 2's in its base-3 representation then add multiply by 7 and add 3.
    \item Take the previous number, convert it to its binary representation using the smallest power of 2 that is greater than or equal to the number of bits required to represent it, treating it as an unsigned integer. Then, invert each bit, and output the result as an unsigned integer.
    \item Given the previous number, output the smallest number bigger than the previous answer that is a prime number.
    \item Take the previous number and output the first perfect square that comes after it.
    \item Take the previous number and treat its digits as coefficients of a polynomial(MSB corresponds to largest power, LSB to the constant term). Calculate and output p(2).
    \item Take the previous number, take its absolute value, and replace each digit with the corresponding letter in the lower cap alphabet and then upper cap, and 0 becomes the letter x(405->dDxXeE), and output the resulted string.
    \item Take the previous number, take its absolute value, and replace each digit with the first and last letters of its name in upper cap(103->OEZOTE), and output the resulted string.
    \item Take the previous number - let's call it num, calculate num mod 118, the result is a number i. Output the name of the i-th element in the periodic table(not 0-indexed).
    \item Take the previous number: n, and write (n mod 10,000) in Roman numerals.
    \item Take the previous number n, and output the day of the week with index (n mod 7) - Sunday is the first day which its index is 0, in lower caps.
    \item Take the previous string, change the letters' caps to alternating caps starting with lower caps, and then reverse and output the resulted string.
    \item Sort the previous string by ASCII order.
    \item Take the previous string and subtract 1 to each character, a becomes z.
    \item Take the previous string, replace each second letter in the string with the next letter in the alphabet and output the resulted string.
    \item Take the previous string, and move its characters such that each letter(regardless of lower/upper caps) that is before 'm' in alphabetical order should move to the start, and every letter after should move to the end. The order must remain - it should be stable.
    \item Take the previous string and surround it with "abcde", "edcba": "abcde" is before and "edcba" is after.
    \item Take the previous string - let's call it s, and output the concatenation of s with itself two times s+s+s.
    \item Take the previous string, encrypt it using Caesar Cipher with key=8 and output the result.
    \item Take the previous string and replace each vowel character(a,A,e,E,i,I,o,O,u,U) with "gh".
    \item Take the previous string, find the longest prefix of the string where characters appear in non-decreasing lexicographic order, and move this prefix to the end of the string and output the resulted string.
    \item Take the previous string and output the sum of the alphabetical positions of only its alphabetical characters (A=1, B=2, ..., Z=26), upper/lower caps don't matter
    \item Take the previous string and output the sum of ASCII values of the string.
\end{enumerate}
\FloatBarrier

There is also a code version of each of the NL instructions, in each of the programming languages supported by PACIFIC. Either NL or code instructions can be used by PACIFIC.

\subsection{Prompt and Parsing Pattern}
\label{append:prompt-pattern}

This section details the prompts employed in both \textit{Prompt Mode} and \textit{Chat Mode}, as well as the parsing strategy used to extract model outputs.

\subsubsection{Prompt Mode.}
\begin{quote}
    You are given an input to an initial instruction, and a list of instructions coded in \{prog\_lang\}. Follow all the instructions, and output a literal (no unsimplified expressions, no function calls) containing the output when executing the provided code on the given input and then following the instructions, even if the function is incorrect or incomplete. You can assume every library/namespace is included/imported. Do NOT output any extra information. Execute the program and the instructions step by step before arriving at an answer. For each instruction with number i, provide the correct output inside [ANSWER][i] and [\textbackslash ANSWER] tags, such that for example the answer for the fourth instruction should be inside [ANSWER][4]  [\textbackslash ANSWER]. You MUST NOT use the [ANSWER] tags anywhere else beside the answers. You can use only one pair of tags per instruction, so you should use exactly \{n\_instructions\} pairs of [ANSWER][i] [\textbackslash ANSWER] tags.

    Input:
    \{input\_string\}
    
    List of instructions:
    
    \{inst\_string\}
    
    Output the result of each instruction inside the [ANSWER] [\textbackslash ANSWER] tags.
\end{quote}

\subsubsection{Chat Mode.}
\begin{quote} 
    You are given an input to an initial instruction, and the you are given different instructions coded in \{prog\_lang\} one by one(one instructions per message) and you will need to follow all of them. Follow all the instructions, and output a literal (no unsimplified expressions, no function calls) containing the output when executing the provided code on the given input and then following the instructions, even if the function is incorrect or incomplete. You can assume every library/namespace is included/imported. Do NOT output any extra information. Execute the program and the instructions step by step before arriving at an answer. For each instruction with number i, provide the correct output inside [ANSWER][i] and [\textbackslash ANSWER] tags, such that for example the answer for the fourth instruction should be inside [ANSWER][4]  [\textbackslash ANSWER]. You MUST NOT use the [ANSWER] tags anywhere else beside the answers.

    Input:
    \{input\_string\}
    
    \{first\_instruction\}
\end{quote}

\subsubsection{Parsing Pattern.}

To ensure deterministic evaluation, we employ a rule-based parsing strategy. For the $i$-th instruction, the expected output is extracted using the following pattern:

\begin{quote}
[ANSWER][i] \textit{<parsed answer>} [\textbackslash ANSWER]
\end{quote}

\subsection{Per-Model Performance Details} \label{append:modelPerformanceFigures}
This section provides detailed performance charts for each model across all 15 benchmark configurations. Each figure includes two subplots corresponding to the two evaluation metrics.

\subsubsection{Claude-3.5-Haiku}
\begin{figure*}[htbp]
    \centering
    \includegraphics[width=\linewidth]{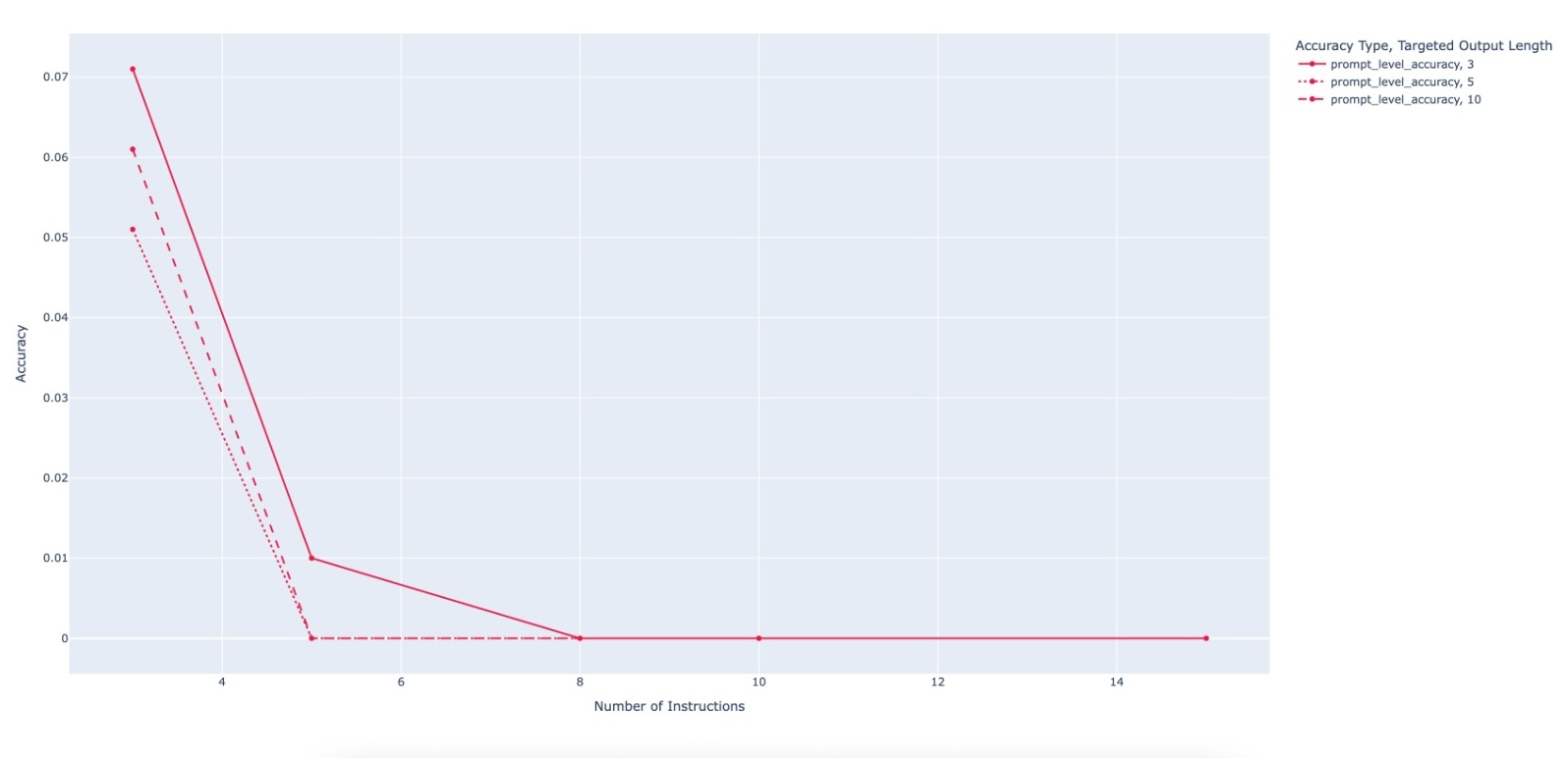}
    \caption{\textbf{Claude-3.5-Haiku Performance.} Results across all benchmarks for Metric Prompt Level Accuracy.}
    \label{fig:claude_35_haiku}
    \Description{Claude-3.5-Haiku Performance}
\end{figure*}
\subsubsection{Claude-4-Sonnet}
\begin{figure*}[htbp]
    \centering
    \includegraphics[width=\linewidth]{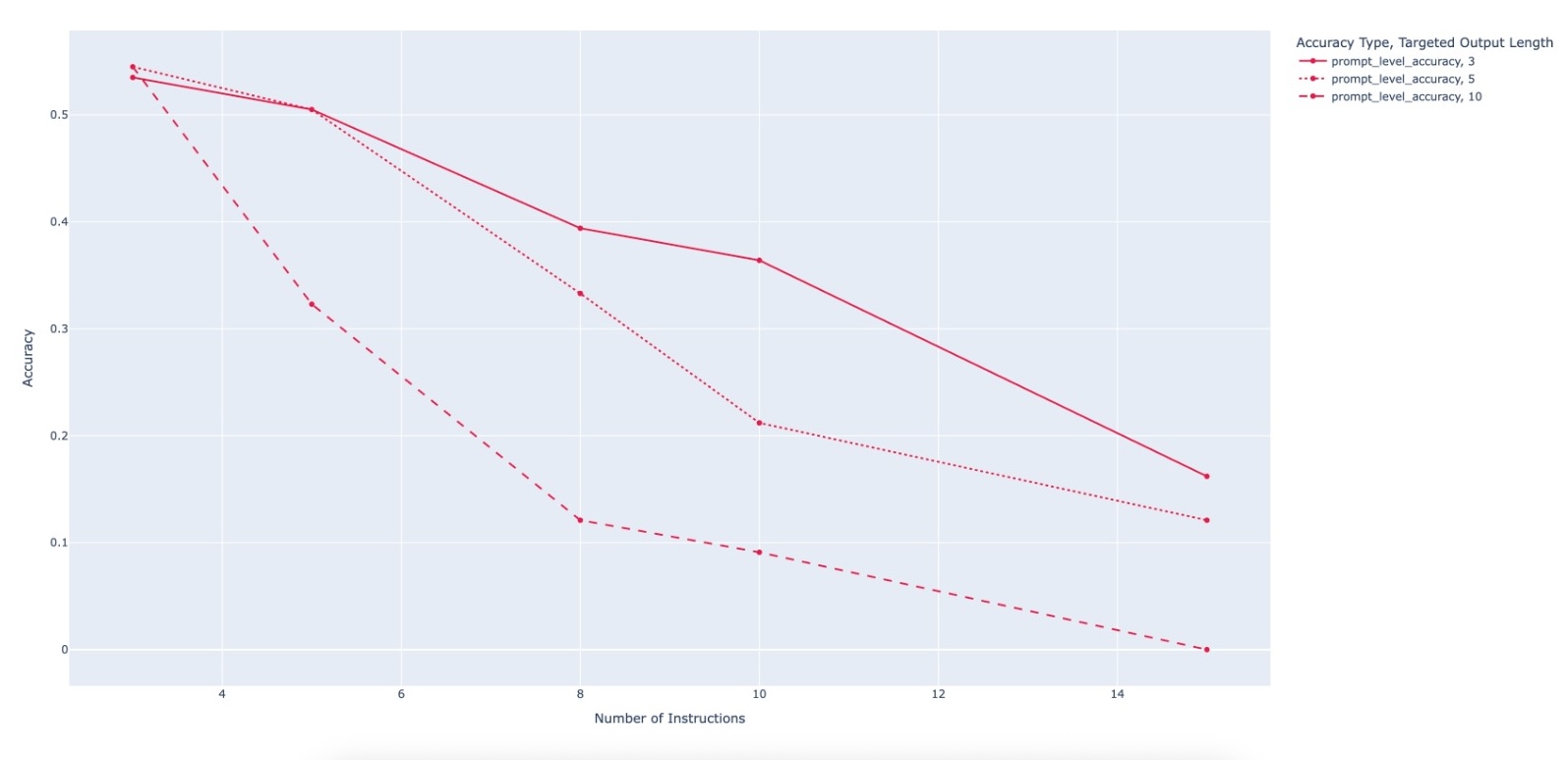}
    \caption{\textbf{Claude-4-Sonnet Performance.} Results across all benchmarks for Metric Prompt Level Accuracy.}
    \label{fig:claude_4_sonnet}
    \Description{Claude-4-Sonnet Performance}
\end{figure*}
\subsubsection{GPT-OSS-120B}
\begin{figure*}[htbp]
    \centering
    \includegraphics[width=\linewidth]{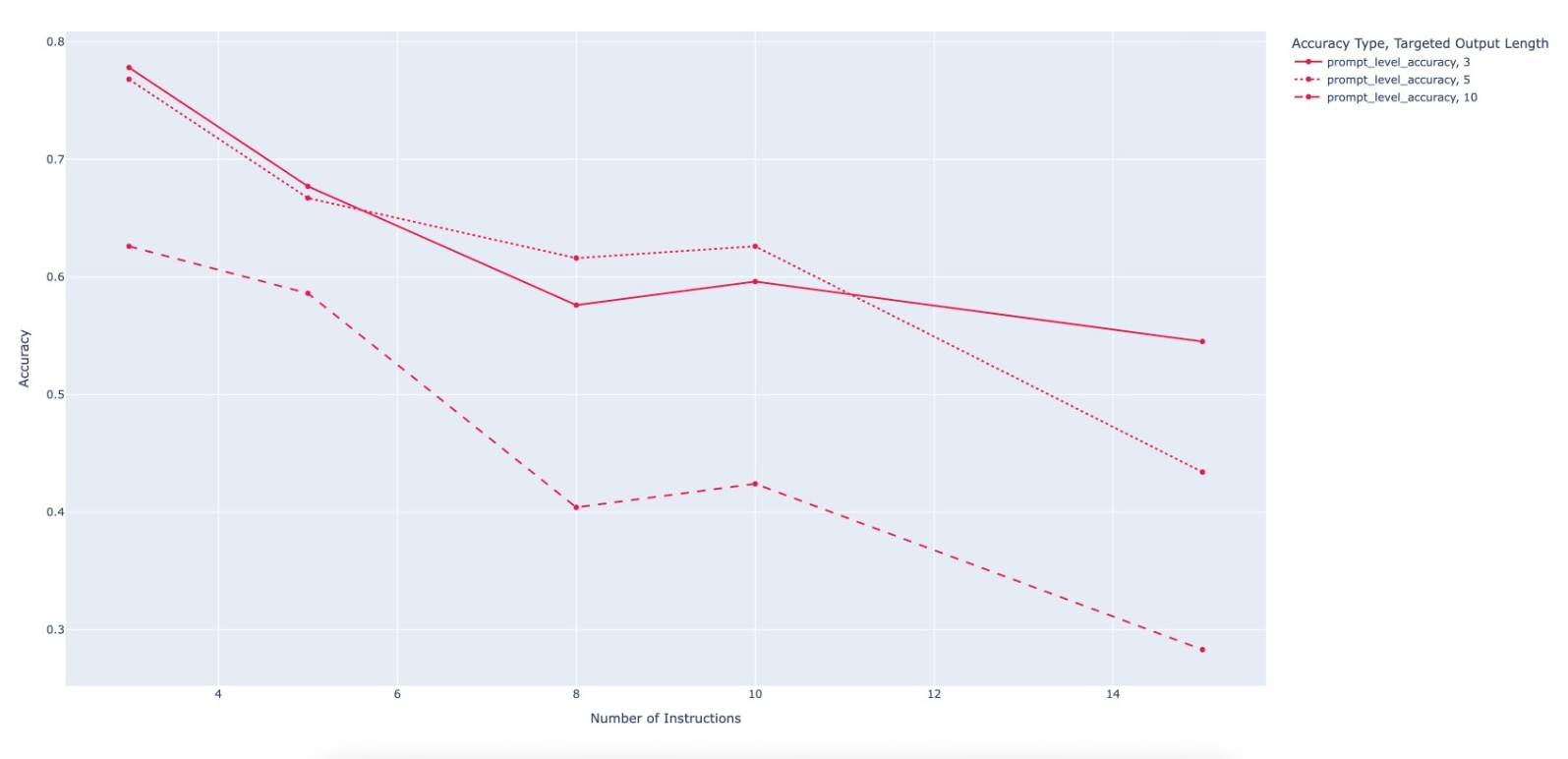}
    \caption{\textbf{GPT-OSS-120B Performance.} Results across all benchmarks for Metric Prompt Level Accuracy.}
    \label{fig:gpt_oss_120b}
    \Description{GPT-OSS-120B Performance}
\end{figure*}
\subsubsection{Llama-4-Maverick-17B-128E-Instruct}
\begin{figure*}[htbp]
    \centering
    \includegraphics[width=\linewidth]{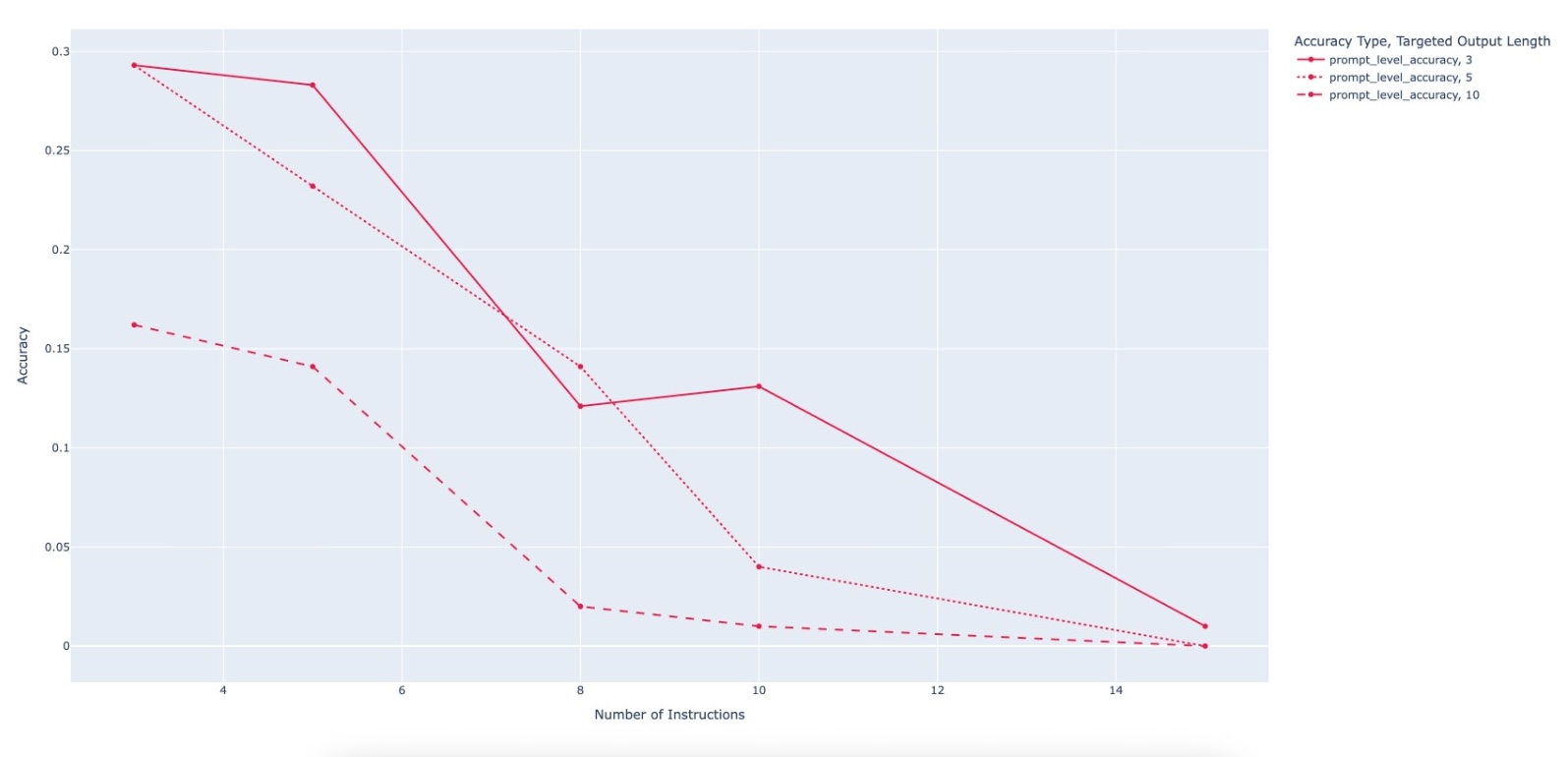}
    \caption{\textbf{Llama-4-Maverick-17B-128E-Instruct Performance.} Results across all benchmarks for Metric Prompt Level Accuracy.}
    \label{fig:llama_4_maverick}
    \Description{Llama-4-Maverick-17B-128E-Instruct Performance}
\end{figure*}
\subsubsection{Qwen3-235B-A22B-Instruct}
\begin{figure*}[htbp]
    \centering
    \includegraphics[width=\linewidth]{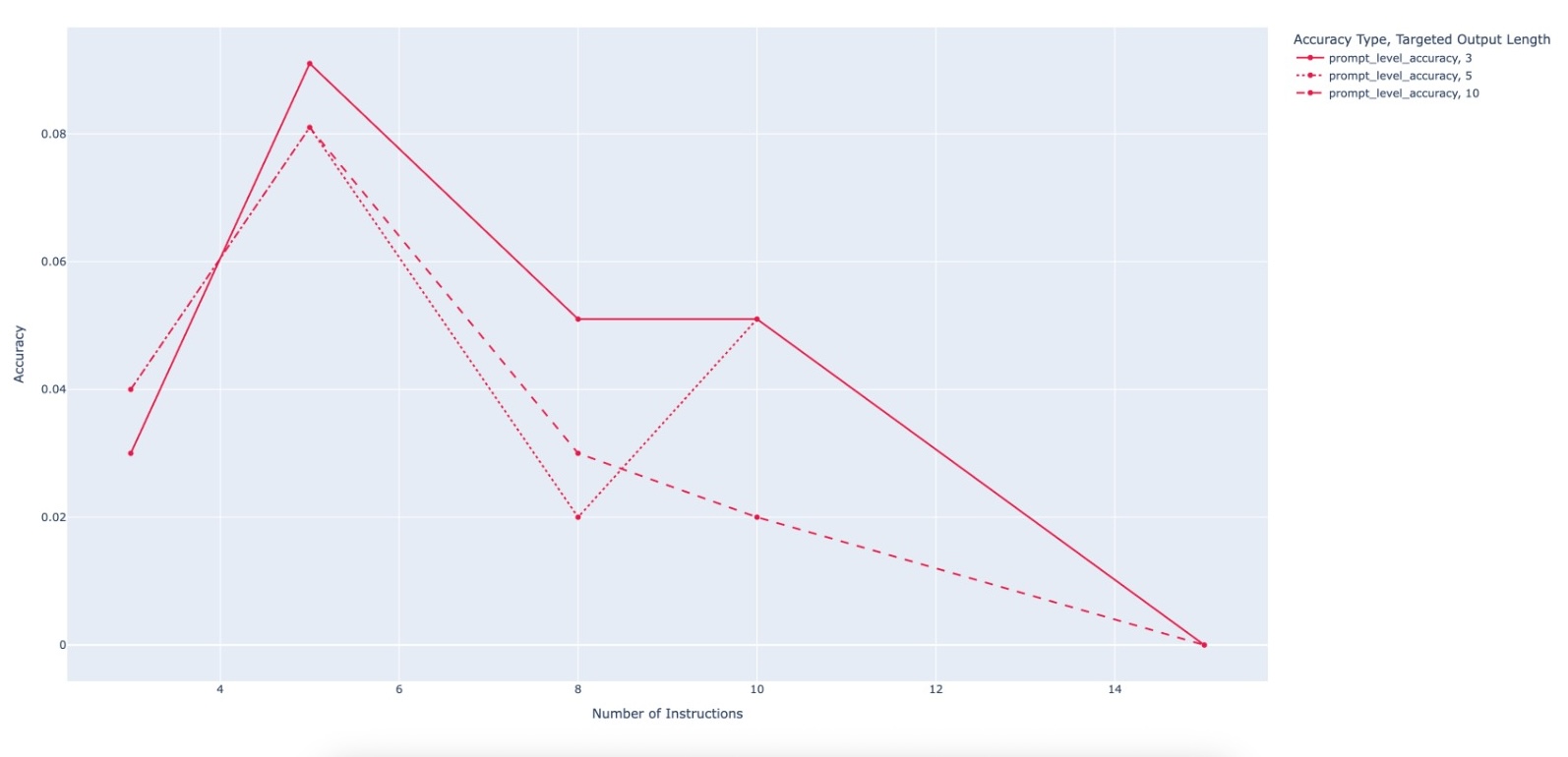}
    \caption{\textbf{Qwen3-235B-A22B-Instruct Performance.} Results across all benchmarks for Metric Prompt Level Accuracy.}
    \label{fig:qwen3_235b_instruct}
    \Description{Qwen3-235B-A22B-Instruct Performance}
\end{figure*}
\subsubsection{Qwen3-Coder-480B-A35B-Instruct}
\begin{figure*}[htbp]
    \centering
    \includegraphics[width=\linewidth]{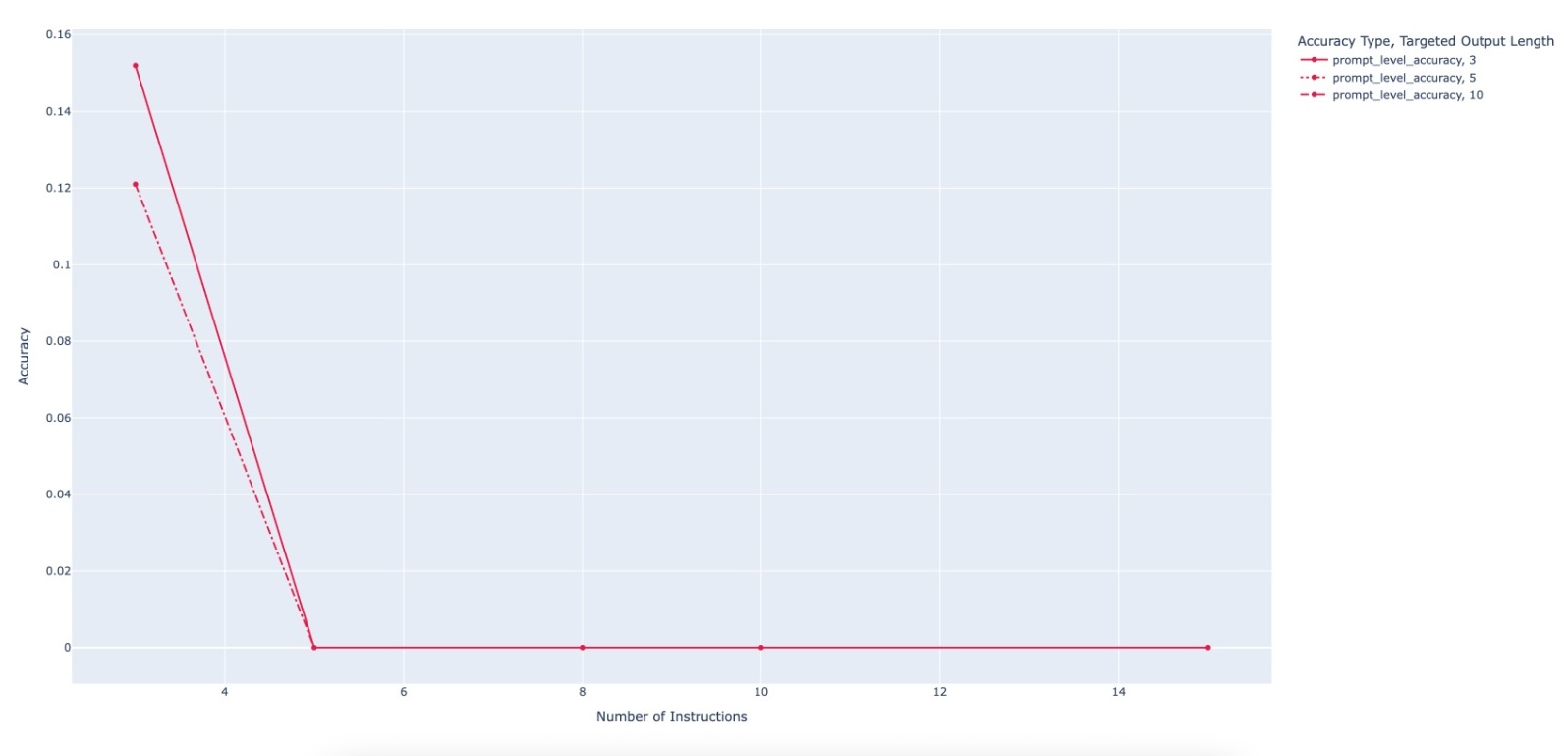}
    \caption{\textbf{Qwen3-Coder-480B-A35B-Instruct Performance.} Results across all benchmarks for Metric Prompt Level Accuracy.}
    \label{fig:qwen_coder_480b}
    \Description{Qwen3-Coder-480B-A35B-Instruct Performance}
\end{figure*}

\end{document}